\documentclass{aa}    

\usepackage{graphicx}
\usepackage{txfonts}

\usepackage{natbib}
\usepackage{dcolumn,longtable}

\newcolumntype{d}{D{.}{.}{-1}}

\begin{document} 

  \title{Inversion of asteroid photometry from Gaia DR2 and the Lowell Observatory photometric database}             
    
  \titlerunning{Inversion of Gaia DR2 and Lowell Observatory asteroid photometry}

  \author{J. \v{D}urech         \inst{1}        \and
          J. Hanu\v{s}          \inst{1}        \and
          R. Van\v{c}o          \inst{1,2}      
         }

  \institute{Astronomical Institute, Faculty of Mathematics and Physics, Charles University, V Hole\v{s}ovi\v{c}k\'ach~2, 180\,00 Prague~8, Czech Republic\\
             \email{durech@sirrah.troja.mff.cuni.cz}
             \and
             Czech National Team, N\'adra\v{z}\'i 549, \'Ujezd u Brna, 664 53, Czech Republic 
             }

  \date{Received ?; accepted ?}

  \abstract
  {Rotation properties (spin-axis direction and rotation period) and coarse shape models of asteroids can be reconstructed from their disk-integrated brightness when measured from various viewing geometries. These physical properties are essential for creating a global picture of structure and dynamical evolution of the main belt.}
  {The number of shape and spin models can be increased not only when new data are available, but also by combining independent data sets and inverting them together. Our aim was to derive new asteroid models by processing readily available photometry.}
  {We used asteroid photometry compiled in the Lowell Observatory photometry database with photometry from the Gaia Data Release 2. Both data sources are available for about 5400 asteroids. In the framework of the Asteroids@home distributed computing project, we applied the light curve inversion method to each asteroid to find its convex shape model and spin state that fits the observed photometry.}
  {Due to the limited number of Gaia DR2 data points and poor photometric accuracy of Lowell data, we were able to derive unique models for only $\sim$\,1100 asteroids. Nevertheless, 762 of these are new models that significantly enlarge the current database of about 1600 asteroid models.}
  {Our results demonstrate the importance of a combined approach to inversion of asteroid photometry. While our models in general agree with those obtained by separate inversion of Lowell and Gaia data, the combined inversion is more robust, model parameters are more constrained, and unique models can be reconstructed in many cases when individual data sets alone are not sufficient.}


  \maketitle

  \section{Introduction}

    The ESA Gaia mission \citep{Gaia:16} was launched in December 2013 with the main aim being to provide accurate astrometry and photometry for more than a billion stars. Gaia systematically scans the whole sky and observes not only stars but also other point-like sources -- asteroids. The second data release (DR2) from April 2018 \citep{Gaia:18} contains astrometric and photometric data for about 14,000 asteroids \citep{Spo.ea:18}. The first analysis of Gaia asteroid photometry was published by \cite{Mom.ea:18}, who performed a population-level study of asteroid shapes and pole orientations. Inversion of photometry of individual asteroids was done by \cite{Dur.Han:18}. Because the number of Gaia DR2 measurements for individual asteroids is small, a successful inversion was possible only for a small fraction of asteroids released in DR2. In \cite{Dur.Han:18}, we reconstructed spins and shapes of 173 asteroids.

    The purpose of this paper is to further exploit DR2 asteroid photometry by combining it with complementary ground-based photometric data.
    Ideally, to maximize the scientific output of the information content of the data, all available photometry should be processed together, but because photometric data are scattered in many individual sources, this is hard to do efficiently in practice. To take the first step, we combined what was readily available -- Lowell and Gaia data. We present new shape and spin models and make them available on the Database of Asteroid Models from Inversion Techniques \citep[DAMIT,][]{Dur.ea:10}.\footnote{\url{http://astro.troja.mff.cuni.cz/projects/asteroids3D}} 
  
  \section{Inversion of Gaia and Lowell photometry}
    
    We used a similar approach as in our previous works: \cite{Dur.ea:07, Dur.ea:09, Dur.ea:16, Dur.ea:18c} and \cite{Han.ea:11, Han.ea:13b}. Namely, we used the light curve inversion method of \cite{Kaa.ea:01} to reconstruct the rotation parameters (rotation period $P$ and the direction of the rotation axis in ecliptic coordinates $\lambda, \beta$) and convex shapes of asteroids. The method and its implementation are described in \cite{Han.ea:11}. In short, the best model is defined as that with the best goodness of fit measured by $\chi^2$. The global minimum in $\chi^2$ is found by scanning the relevant parameter space densely with gradient-based optimization of \cite{Kaa.ea:01}. In practice, this is done with Asteroids@home distributed computing project \citep{Dur.ea:15}.

    We used two sources of asteroid photometry: the Lowell Observatory photometry database, which contains re-processed photometry from the largest ground-based surveys reported to the Minor Planet Center \citep{Osz.ea:11, Bow.ea:14}, and Gaia DR2 \citep{Spo.ea:18}. While both sources contain asteroid disk-integrated photometry, their precision and cadence are very different.

    Lowell data are available for approximately 330\,000 asteroids, and they cover more than a decade from 1998 to 2011. Magnitudes are given for V filter, and typically there are hundreds of measurements for a single object. What makes this data set problematic and not very efficient regarding inversion is low accuracy of photometry, which is typically only 0.1--0.2\,mag. On the contrary, Gaia data (magnitudes in G band) are very accurate with errors of 0.01\,mag. Their disadvantage is that they cover only a limited time interval of 22 months in 2014--16 and the number of measurements per asteroid is low. From 14,000 asteroids reported in DR2, only about 5500 have at least ten observations. Therefore, separate inversion of either Lowell or Gaia data alone is not very efficient. In our previous publications, we derived 328 new models from Lowell data \citep{Dur.ea:16} and 129 new models from Gaia DR2 \citep{Dur.Han:18}. However, it is straightforward to combine both data sets and invert them together. For example, we successfully combined Lowell data with WISE thermal observations in \cite{Dur.ea:18c}. 

    We selected all asteroids with photometric measurements in the Lowell Observatory database and in Gaia DR2. We required that the number of Gaia observations was greater than or equal to ten. Together, the combined data set consisted of 5412 asteroids.
    
    \subsection{Weights and thresholds}

      Because Lowell and Gaia data have very different photometric accuracy, it is necessary to weight them differently when computing $\chi^2$. Formally, $\chi^2$ is computed as a sum of squares of residuals between model and data divided by measurement errors. However, reliable errors of individual measurements are not known and are often dominated by errors other than Gaussian random noise. Therefore, we minimize a relative goodness-of-fit value
      \begin{equation}
       \chi^2_\text{rel} = \sum_\text{Gaia} \left(L_i^\text{obs} - L_i^\text{model}\right)^2 + w\,\sum_\text{Lowell} \left(L_j^\text{obs} - L_j^\text{model}\right)^2\,,
      \end{equation}
      where the total $\chi^2$ is composed of two parts weighted by $w$. If we assume that a typical uncertainty of Lowell photometry is 0.1\,mag and that of Gaia is 0.01\,mag, the weight of Lowell data with respect to Gaia should be $w = 0.01$.  To check the sensitivity of our results to the particular choice of $w$, we also tested values $w = 0.05$ and 0.1 (Table~\ref{tab:weights}).

      The best-fit model is defined as having the lowest $\chi^2_\text{rel}$. The period parameter space is typically full of other local minima. We accept the global minimum $\chi^2_\text{min}$ as a unique solution if all other models with different parameters have $\chi^2$ values higher than $\chi^2_\text{tr} = (1 + k \sqrt{2/\nu})\,\chi^2_\text{min}$, where $\nu$ is the number of degrees of freedom and $k$ is selected such that it produces the maximum number of correct solutions while giving a negligible number of false positive solutions. The smaller $k$, the closer $\chi^2_\text{tr}$ is to $\chi^2_\text{min}$ and the more models are accepted as reliable. At the same time, small $k$ increases the number of false positive solutions. If $k$ is large, $\chi^2_\text{tr}$ is significantly larger than $\chi^2_\text{min}$ and the global minimum has to be deep with respect to other minima to be accepted as a unique solution of the inverse problem. Thus, large $k$ minimizes the number of false positive solutions but also the number of accepted models. The value of $k$ is set based on a trade-off between these two trends. For example, in \cite{Dur.ea:18c} we used $k = 1/2$, in \cite{Dur.Han:18} we used $k = 1$.

      To test which values of $k$ and $w$ are best for our case of inversion of Lowell and Gaia data, we tried three values of $k = 1, 1/2, 1/3$ and $w = 0.01, 0.05, 0.1$. The results are given in Table~\ref{tab:weights}. For different values of $k$ and $w$, we counted the total number of accepted models. We then selected those accepted models that had reliable (uncertainty code $\mathrm{U} \geq 3$) periods listed in the Light curve Database (LCDB) of \cite{War.ea:09}, version from November 20, 2018, and compared that period with the value we obtained from inversion of Lowell and Gaia data. We assumed that our model was correct if the relative difference between periods was less than 10\%. The relative fraction of correctly determined periods is listed in Table~\ref{tab:weights}. For the final analysis, we adopted values $w = 0.01$, $k = 3$ for convex models and $k = 1$ for ellipsoids. The fraction of incorrect periods is not negligible ($\sim 15\%$) but most of these incorrect periods do not pass further reliability checks, which include comparison between periodograms based on convex models (with different resolution) and ellipsoids, visual verification of the data fits, rotation around the shortest axis, and so on \citep[for more details see][Fig.~3]{Dur.ea:18c}.

      \begin{table}
       \centering
       \begin{tabular}[h]{c|ccc|r}
                & \multicolumn{3}{c|}{convex models}            & \multicolumn{1}{c}{ellipsoids}         \\
        $k$     & $w = 0.01$    & $w = 0.05$    & $w = 0.1$     & \multicolumn{1}{c}{$w = 0.01$}        \\
        \hline
        1       & 287 \ 87\%    & 198 \ 92\%    & 151 \ 95\%    & 710 \ 90\%    \\
        1/2     & 522 \ 86\%    & 403 \ 86\%    & 326 \ 88\%    & 997 \ 85\%    \\
        1/3     & 641 \ 83\%    & 513 \ 82\%    & 453 \ 82\%    & 1215 \ 78\%    \\
        \hline
       \end{tabular}
       \caption{Comparison between our best-fit periods and LCDB periods for different setups of $k$ and $w$. Each cell of the table contains the number of models with uniquely determined period and the fraction (in percent) of those models for which the period was correct, i.e., the same (within $\pm 10\%$) as that in the LCDB.}
       \label{tab:weights}              
      \end{table}

    \subsection{Comparison with Gaia-only models}
    \label{sec:models_comparison}

      In \cite{Dur.Han:18}, we derived rotation periods and pole directions for 129 asteroids from Gaia DR2 data. All these asteroids were included also in our new Gaia and Lowell combined data set so that we could compare results based on only Gaia photometry with those based on combined photometry. From the 129 models, 74 were in agreement -- having the same rotation period and pole direction (within the uncertainty intervals). Only two models were completely different: (1540)~Kevola and (15496)~1999~DQ3. For Kevola, Gaia-only data produced a model that was incorrect. With the combined data set, the new rotation period of 20.08\,hr is consistent with the period given in the LCDB. For asteroid (15496), the new rotation period of 88.295\,h was one of local minima in the Gaia-based periodogram, the global minimum was at 81.21\,h.

      In another eight cases, rotation periods were the same but pole directions were different -- typically prograde or retrograde differences -- which was likely because Gaia data were not sufficient to determine the pole correctly.

      In the remaining 45 cases, we did not retrieve any unique model based on combined photometry because the best-fit models did not pass reliability checks. The reasons were: a non-unique period in 17 cases, more than two pole solutions in 20 cases, and an incorrect inertia tensor in 8 cases.

    \subsection{Comparison with DAMIT models}  
 
      After processing all of the approximately 5500 asteroids for which we had photometric data from both sources, we obtained models for 1142 asteroids that passed all tests. Out of these, 367 were already in DAMIT. Similarly as in \cite{Dur.ea:16, Dur.ea:18c} we compared spin solutions of our new models with those in DAMIT. Although DAMIT models cannot be treated as completely independent because in many cases they were reconstructed also from Lowell data \citep{Dur.ea:16, Dur.ea:18c}, they can be to some extent used for consistency checks. The agreement with DAMIT models was good, most of the asteroids had the same spin solution within the expected uncertainty intervals. There were only four clearly incorrect solutions with completely different periods and 11 cases for which the periods were very similar (relative difference $\sim 10^{-4}$) but they corresponded to different local minima and thus the poles were significantly different ($> 60^\circ$ apart). However, it was not clear which model was correct because in many of these 11 cases, the DAMIT model was based on sparse data only \citep{Han.ea:16} or combination of Lowell and WISE data \citep{Dur.ea:18c}. In another few cases, the formal discrepancy between pole directions of new and DAMIT models was caused by the fact that they were mirror shape solutions with a difference in the pole ecliptic longitude of  $\lambda \pm 180^\circ$.

      When comparing our new results with DAMIT, we realized that four DAMIT models were incorrect. The updated models reconstructed from all available light curves and sparse photometry are listed in Table~\ref{tab:updated_models}.

      \paragraph{(208) Lacrimosa} The new model has prograde rotation while previous studies published retrograde models \citep{Sli.ea:03, Dur.ea:11}. This was caused by the slightly incorrect sidereal period of the published solution. 

      \paragraph{(526) Jena} The model published by \cite{Han.ea:16} lists an incorrect rotation period of 9.52\,h because the best-fit period was searched only near the value of 9.474\,h reported by \cite{Bar.ea:94}. The new model with a period 11.88\,h is clearly better and fits well not only the light curves of \cite{Bar.ea:94} but also sparse USNO and Gaia photometry.

      \paragraph{(1332) Marconia} Model in DAMIT published by \cite{Dur.ea:16} was reconstructed from only Lowell Observatory data and its rotation period was 19.2\,h. According to our new analysis of Lowell and Gaia data, the correct rotation period is 32.1\,h, which agrees with the value published by \cite{Dev.ea:17}. 

      \paragraph{(4800) Veveri} The rotation period determined by \cite{Han.ea:18a} of 7.14\,h is incorrect. Our new analysis of light curves from \cite{Han.ea:18a}, Lowell, and Gaia photometry gives a rotation period of 6.21\,h.

      \begin{table*}
       \centering
       \begin{tabular}[t]{r l r r r r d l}
        \multicolumn{2}{c}{Asteroid}            & \multicolumn{1}{c}{$\lambda_1$}       & \multicolumn{1}{c}{$\beta_1$}   & \multicolumn{1}{c}{$\lambda_2$}       & \multicolumn{1}{c}{$\beta_2$}   & \multicolumn{1}{c}{$P$}       & Light curves   \\
        Number  & Name  & \multicolumn{1}{c}{[deg]}             & \multicolumn{1}{c}{[deg]}     & \multicolumn{1}{c}{[deg]}               & \multicolumn{1}{c}{[deg]}     & \multicolumn{1}{c}{[h]} &               \\
        \hline
        208     & Lacrimosa     & 20    & 43    & 198   & 50    & 14.08568      & \cite{Bin:87}, \cite{Sli.Bin:96},       \\                              
                &               &       &       &       &       &               & \cite{Ste:14}, USNO, Gaia DR2           \\
        526     & Jena          & 16    & 50    & 194   & 54    & 11.87655      & \cite{Bar.ea:94}, USNO, Gaia DR2        \\
        1332    & Marconia      & 59    & 26    & 240   & 26    & 32.11995      & \cite{Ste:13}, \cite{Dev.ea:17},        \\
                &               &       &       &       &       &               & USNO, Gaia DR2 \\
        4800    & Veveri        & 94    & $-9$  & 274   & 16    & 6.214858      & \cite{Han.ea:18a}, Lowell, Gaia DR2     \\
        \hline
       \end{tabular}
       \caption{Updated spin states of incorrect DAMIT models. For each asteroid, two possible pole directions in the ecliptic coordinates $(\lambda, \beta)$ and the sidereal rotation period $P$ are given. Sources of light curves and sparse photometry used for model reconstruction are listed in the last column.}
       \label{tab:updated_models}               
      \end{table*}

    \subsection{New spin and shape models}
    \label{sec:new_models}

    In total, we derived new models for 775 asteroids. Many of them had their rotation period listed in the LCDB and in most cases the LCDB period agreed with our value. However, we identified 23 asteroids for which we obtained a model with a period that was clearly different from the value in the LCDB. We checked the original sources of LCDB records for all discrepant cases with uncertainty code $\geq 2+$. In 13 cases, we realized that our model was incorrect because the LCDB period was reliable and did not agree with our solution. We removed these asteroids from the list of new models. In some cases, however, even for $U = 3$, the LCDB period seemed not robust enough to reject our solution. The spin parameters of our models, LCDB period values with possible notes, and the number of data points we used for inversion are listed in Table~A.1. Shape models and photometric data are available also through DAMIT. 

    We also derived simple single-period models for asteroids that are in fact (according to independent photometric observations) in a non-principal axis rotation state. Apparently, our principal-axis-rotation model is good enough given the quality and coverage of the data to fit the main period in light curves of some tumblers. For example, our principal-axis-rotation model of (5247)~Krylov has the rotation period of 82.291\,h, which is the most prominent period of its light curves given by a combination of two physical periods of free precession: $1/82.291 \approx 1/68.15\,\text{h} - 1 / 396.30\,\text{h}$ \citep{Lee.ea:17, Lee.ea:18}.

  \section{Conclusion}  

    The set of 762 new asteroid models that we reconstructed from combined inversion of Lowell and Gaia DR2 photometry significantly increases the total number of asteroids with known spin state and shape. Although the sample of models in Table~A.1 may contain erroneous models, their fraction should not be higher than a few percent. The results we publish can be used for the analysis of the distribution of spins and shapes of asteroids in the main belt, collisional families, and other sub-populations. However, detailed models of individual asteroids require dense light curves observed under various geometries that better constrain the shape and spin parameters. 
      
  \begin{acknowledgements}
    The authors were supported by the grant 18-04514J of the Czech Science Foundation. We greatly appreciate contribution of tens of thousand of volunteers who joined the Asteroids@home BOINC project and provided their computing resources. This work has made use of data from the European Space Agency (ESA) mission {\it Gaia} (\url{https://www.cosmos.esa.int/gaia}), processed by the {\it Gaia} Data Processing and Analysis Consortium (DPAC, \url{https://www.cosmos.esa.int/web/gaia/dpac/consortium}). Funding for the DPAC has been provided by national institutions, in particular the institutions participating in the {\it Gaia} Multilateral Agreement.

  \end{acknowledgements}

\newcommand{\SortNoop}[1]{}

  \onecolumn
  \begin{appendix}
    \section{List of new models}

    \small
\begin{longtable}{r l r r r r @{} d @{} d l c r r c}
      \caption{\label{tab:models} List of new asteroid models. For each asteroid, we list one or two pole directions in the ecliptic coordinates $(\lambda, \beta)$, the sidereal rotation period $P$, the rotation period $P_\mathrm{LCDB}$ from the LCDB and its quality code $U$ (if available), the number of sparse photometric data points $N_\mathrm{L}$ and $N_\mathrm{G}$ in the Lowell database and Gaia DR2, respectively, and the method which was used to derive the rotation period: C -- convex models, E -- ellipsoids, CE -- both methods gave the same unique period. The accuracy of the sidereal rotation period $P$ is of the order of the last decimal place given. For some asteroids, the LCDB contains a note about the rotation state or binarity: Here A means an ambiguous period; B stands for a binary system; T means tumbling -- a non-principal-axis rotation; T0 means that the theoretical tumbling damping time-scale \citep{Pra.ea:14} is long enough that tumbling might be expected, but observations are not sufficient to substantiate either tumbling or not tumbling; T$-$ means that the tumbling damping time scale is long enough that tumbling might be expected, but observations indicate that the object is not tumbling; and T+ indicates that the tumbling damping time scale is short enough that tumbling would not seem likely, however observations indicate that it may be tumbling or indeed is tumbling.}\\
      \hline\hline
      \multicolumn{2}{c}{Asteroid}              & \multicolumn{1}{c}{$\lambda_1$}       & \multicolumn{1}{c}{$\beta_1$}   & \multicolumn{1}{c}{$\lambda_2$}       & \multicolumn{1}{c}{$\beta_2$}   & \multicolumn{1}{c}{$P$}       & \multicolumn{1}{c}{$P_\mathrm{LCDB}$} & \multicolumn{1}{c}{U}   &       Note    & \multicolumn{1}{c}{$N_\mathrm{L}$}    & \multicolumn{1}{c}{$N_\mathrm{G}$}      & method        \\
      number    & name/designation              & \multicolumn{1}{c}{[deg]}             & \multicolumn{1}{c}{[deg]}       & \multicolumn{1}{c}{[deg]}             & \multicolumn{1}{c}{[deg]}       & \multicolumn{1}{c}{[h]}       & \multicolumn{1}{c}{[h]}                 &                       &               &                               &                               &               \\
      \hline
      \endfirsthead
      \caption{continued.}\\
      \hline\hline
      \multicolumn{2}{c}{Asteroid}              & \multicolumn{1}{c}{$\lambda_1$}       & \multicolumn{1}{c}{$\beta_1$}   & \multicolumn{1}{c}{$\lambda_2$}       & \multicolumn{1}{c}{$\beta_2$}   & \multicolumn{1}{c}{$P$}       & \multicolumn{1}{c}{$P_\mathrm{LCDB}$} & \multicolumn{1}{c}{U}   &       Note    & \multicolumn{1}{c}{$N_\mathrm{L}$}    & \multicolumn{1}{c}{$N_\mathrm{G}$}      & method        \\
      number    & name/designation              & \multicolumn{1}{c}{[deg]}             & \multicolumn{1}{c}{[deg]}       & \multicolumn{1}{c}{[deg]}             & \multicolumn{1}{c}{[deg]}       & \multicolumn{1}{c}{[h]}       & \multicolumn{1}{c}{[h]}                 &                       &               &                               &                               &               \\
      \hline
      \endhead
      \hline
      \endfoot
         47 & Aglaja                    &   150 &    17 &   327 &    12 & 13.17710 &      13.178 &     3 &        &   439 &   24 &   E  \\
  191 & Kolga                     &   170 & $-39$ &   328 & $-42$ & 17.60390 &      17.604 &     3 &        &   520 &   26 &  CE  \\
  206 & Hersilia                  &   250 &    43 &       &       & 11.11129 &      11.122 &     3 &        &   502 &   17 &   E  \\
  209 & Dido                      &    43 & $-34$ &   217 & $-39$ &  5.73561 &      5.7366 &     3 &        &   430 &   14 &   C  \\
  217 & Eudora                    &   135 & $-25$ &       &       &  25.2626 &      25.272 &     3 &        &   416 &   29 &   E  \\
  231 & Vindobona                 &   223 &    49 &       &       & 14.24470 &      14.245 &     3 &        &   335 &   15 &   C  \\
  238 & Hypatia                   &    84 &    20 &   261 &    45 &  8.87274 &      8.8749 &     3 &        &   334 &   20 &   E  \\
  244 & Sita                      &    41 & $-80$ &   218 & $-74$ &  129.716 &      129.51 &  3$-$ &    T0  &   376 &   12 &  CE  \\
\end{longtable}

  \end{appendix}

\end{document}